\documentclass[a4paper]{article}

\usepackage{icrc2013}

\title{The ASTRI Mini-Array Science Case}

\shorttitle{The ASTRI Mini-Array Science Case}

\authors{
S.~Vercellone$^{1}$, 
G.~Agnetta$^{1}$, 
L.A.~Antonelli$^{2}$, 
D.~Bastieri$^{3}$, 
G.~Bellassai$^{4}$, 
M.~Belluso$^{4}$,
C.~Bigongiari$^{5a}$,
S.~Billotta$^{4}$, 
B.~Biondo$^{1}$, 
G.~Bonanno$^{4}$, 
G.~Bonnoli$^{6}$, 
P.~Bruno$^{4}$, 
A.~Bulgarelli$^{7}$, 
R.~Canestrari$^{6}$,
M.~Capalbi$^{1}$,
P.~Caraveo$^{8}$, 
A.~Carosi$^{2}$, 
E.~Cascone$^{9}$, 
O.~Catalano$^{1}$, 
M.~Cereda$^{6}$, 
P.~Conconi$^{6}$, 
V.~Conforti$^{7}$, 
G.~Cusumano$^{1}$, 
V.~De~Caprio$^{9}$, 
A.~De~Luca$^{8}$,
A.~Di~Paola$^{2}$,
F.~Di~Pierro$^{5a}$, 
D.~Fantinel$^{10}$, 
M.~Fiorini$^{8}$, 
D.~Fugazza$^{6}$, 
D.~Gardiol$^{5b}$, 
M.~Ghigo$^{6}$, 
F.~Gianotti$^{7}$, 
S.~Giarrusso$^{1}$, 
E.~Giro$^{10}$, 
A.~Grillo$^{4}$, 
D.~Impiombato$^{1}$, 
S.~Incorvaia$^{8}$, 
A.~La~Barbera$^{1}$,
N.~La~Palombara$^{8}$,
V.~La~Parola$^{1}$, 
G.~La Rosa$^{1}$, 
L.~Lessio$^{10}$, 
G.~Leto$^{4}$, 
S.~Lombardi$^{2}$, 
F.~Lucarelli$^{2}$,  
M.C.~Maccarone$^{1}$, 
G.~Malaguti$^{7}$, 
G.~Malaspina$^{6}$, 
V.~Mangano$^{1}$, 
D.~Marano$^{4}$, 
E.~Martinetti$^{4}$, 
R.~Millul$^{6}$, 
T.~Mineo$^{1}$, 
A.~Mist\'{o}$^{6}$, 
C.~Morello$^{5a}$,
G.~Morlino$^{11}$,
M.R.~Panzera$^{6}$, 
G.~Pareschi$^{6}$, 
G.~Rodeghiero$^{10}$, 
P.~Romano$^{1}$, 
F.~Russo$^{1}$, 
B.~Sacco$^{1}$, 
N.~Sartore$^{8}$, 
J.~Schwarz$^{6}$, 
A.~Segreto$^{1}$, 
G.~Sironi$^{6}$, 
G.~Sottile$^{1}$, 
A.~Stamerra$^{5a}$, 
E.~Strazzeri$^{1}$, 
L.~Stringhetti$^{8}$, 
G.~Tagliaferri$^{6}$, 
V.~Testa$^{2}$, 
M.C.~Timpanaro$^{4}$, 
G.~Toso$^{6}$, 
G.~Tosti$^{12}$, 
M.~Trifoglio$^{7}$, 
P.~Vallania$^{5a}$, 
V.~Zitelli$^{13}$
(The ASTRI Collaboration$^{14}$), and F.~Tavecchio$^{6}$
}
\afiliations{
$^{1}$ INAF -- IASF Palermo, Via U. La Malfa 153, I--90146 Palermo, Italy \\
$^{2}$ INAF -- Osservatorio Astronomico di Roma, Via Frascati 33, I--00040 Monte Porzio Catone (RM), Italy \\
$^{3}$ Universit\`{a} di Padova, Dip. Fisica e Astronomia, Via Marzolo 8, I--35131 Padova, Italy\\
$^{4}$ INAF -- Osservatorio Astrofisico di Catania, Via S. Sofia 78, I--95123 Catania, Italy\\
$^{5}$ INAF -- Osservatorio Astrofisico di Torino:  (a) Via P. Giuria 1, I--10125 Torino, Italy; 
           (b) Via Osservatorio 20, I--10025 Pino Torinese (TO), Italy \\
$^{6}$ INAF -- Osservatorio Astronomico di Brera, Via E. Bianchi 46, I--23807 Merate (LC), Italy \\
$^{7}$ INAF -- IASF Bologna, Via P. Gobetti 101, I--40129 Bologna, Italy\\
$^{8}$ INAF -- IASF Milano, Via E. Bassini 15, I--20133 Milano, Italy \\
$^{9}$ INAF -- Osservatorio Astronomico di Capodimonte, Salita Moiariello 16, I--80131 Napoli, Italy\\
$^{10}$ INAF -- Osservatorio Astronomico di Padova, Vicolo dell'Osservatorio 5, I--35122 Padova, Italy \\
$^{11}$ INAF -- Osservatorio Astronomico di Arcetri, Largo Enrico Fermi 5, I--50125 Firenze, Italy \\
$^{12}$ Universit\`{a} di Perugia, Dip. Fisica, Via A. Pascoli, I--06123 Perugia, Italy\\
$^{13}$ INAF -- Osservatorio Astronomico di Bologna, Via Ranzani 1, I--40127 Bologna, Italy\\
$^{14}$ \tt{http://www.brera.inaf.it/astri/}
}

\email{stefano.vercellone@iasf-palermo.inaf.it}

\abstract{ASTRI is a Flagship Project financed by the Italian Ministry of Education, University and Research, and led by INAF, 
the Italian National Institute of Astrophysics. Within this framework, INAF is currently developing an end-to-end 
prototype of a Small Size Telescope  in a dual-mirror configuration (SST--2M) for the Cherenkov Telescope Array (CTA), 
scheduled to start data acquisition in 2014. Although the ASTRI SST-2M prototype is mainly a 
technological demonstrator, it will perform scientific observations of the Crab Nebula, Mrk 421 and Mrk 501 at $E\ge1$\,TeV. 
A remarkable improvement in terms of performance could come from the operation, in 2016, of a SST--2M mini-array, 
composed of a few \mbox{SST--2M} telescopes to be placed at final CTA Southern Site. The SST mini-array will be able to 
study in great detail relatively bright sources (a few $\times 10^{-12}$ erg\,cm$^{-2}$s$^{-1}$ at 10 TeV) with angular resolution of a few 
arcmin and energy resolution of about 10--15\%. 
Thanks to the stereo approach, it will be possible to verify the wide field of view (FoV) performance through the detections of very high energy 
showers with core located at a distance up to 500 m, to compare the mini-array performance with the Monte Carlo 
expectations by means of deep observations of selected targets, and to perform the first CTA science
at the beginning of the mini-array operations.
Prominent sources such as extreme blazars (e.g., 1ES 0229$+$200), nearby well-known BL Lac objects (e.g., MKN~501) and radio-galaxies, 
galactic pulsar wind nebulae (e.g., Crab~Nebula, Vela-X), supernovae remnants (e.g., Vela-junior, RX~J1713.7$-$3946), 
micro-quasars (e.g., LS~5039), and the Galactic Center can be observed in a previously unexplored energy range, in order to 
investigate the electron acceleration and cooling, relativistic and non relativistic shocks, the search for cosmic-ray (CR) 
Pevatrons, the study of the CR propagation, and the impact of the extragalactic background light on the spectra of the 
sources.}


\begin{document}
\maketitle

%%%%%%%%%%%%%%%%%%%%%%%%%%%%%%%%%%%%%%%%%%%%%
%%%%%%%%%%%%%%%%%%%%%%%%%%%%%%%%%%%%%%%%%%%%%
	\section{Introduction}
%%%%%%%%%%%%%%%%%%%%%%%%%%%%%%%%%%%%%%%%%%%%%
%%%%%%%%%%%%%%%%%%%%%%%%%%%%%%%%%%%%%%%%%%%%%
%
ASTRI (``Astrofisica con Specchi a Tecnologia Replicante Italiana") is a flagship project of the Italian Ministry of
Education, University and Research strictly linked to the development of the ambitious Cherenkov Telescope 
Array (CTA,~\cite{bib:2011ExA....32..193A,bib:2013APh....43....3A}). 
Within this framework, INAF is currently developing an end-to-end prototype of the CTA small-size telescope 
in a dual-mirror configuration (SST-2M) to be tested under field conditions at the INAF ``M.C. Fraca\-sto\-ro'' observing
station in Serra La Nave (Mount Etna, Sicily)~\cite{bib:Maccarone_icrc13}, and scheduled to start data acquisition in 2014.
A detailed description of the ASTRI Project, the Prototype structure, mirrors, camera, and scientific performance are provided in 
\cite{bib:Pareschi_icrc13,bib:Pareschi_inprep}, \cite{bib:Canestrari_icrc13}, \cite{bib:Catalano_icrc13}, and 
\cite{bib:Bigongiari_icrc13}, respectively.
Although the ASTRI SST--2M will mainly be a technological prototype, it will perform scientific observations
of the Crab Nebula, MRK~421, and MRK~501. Preliminary cal\-cu\-la\-tions show that in the maximum sensitivity range 
($\ge 1$\, TeV) we can obtain a 5$\sigma$ detection of a source at a flux level of 1 Crab in a few hours,
while in the energy range $\ge 10$\,TeV a flux level of 1~Crab at 5$\sigma$ can be reached in a few tens of hours. 
Because of strong flux and spectral variations of the two Markarian sources, estimates of exposures
are more uncertain. In case of large flares, with fluxes up to 5--10 Crab Units (see, e.g.,~\cite{bib:Cortina_ATel4976}), 
detection could be reached on a much shorter time-scale.

%%%%%%%%%%%%%%%%%%%%%%%%%%%%%%%%%%%%%%%%%%%%%
%%%%%%%%%%%%%%%%%%%%%%%%%%%%%%%%%%%%%%%%%%%%%
	\section{The mini-array in a nutshell}
%%%%%%%%%%%%%%%%%%%%%%%%%%%%%%%%%%%%%%%%%%%%%
%%%%%%%%%%%%%%%%%%%%%%%%%%%%%%%%%%%%%%%%%%%%%
%
A remarkable improvement in terms of performance -and expected scientific results- could come from the operation
in 2016 of a mini-array, composed of a few SST--2M telescopes and to be placed at the final CTA Southern 
Site. This could constitute the first {\it seed} of the future CTA Project.
The SST-2M mini-array will verify the following array properties: 
\begin{itemize}
\item the array performance in terms of reliability and cost at the chosen site;
\item the trigger algorithms;
\item the wide field of view performance to detect very high energy showers with the core located at a distance up to 500\,m;
\item the hardware/software configurations for the array;
\item the data-handling chain.
\end{itemize}
Moreover, through deep observations of a few selected targets, it will allow us to compare
the actual performance with the Monte Carlo expectations, and to perform the first CTA science, 
by means of  a few solid detections during the first year.

A preliminary set of tests on the scientific performance of the SST--2M mini-array (with different geometrical
configurations and under a few simplifying assumptions) is reported in~\cite{bib:ASTRI-MC-IFSITO-5000-005}.
These Monte Carlo simulations allowed us to estimate that the mini-array will be able to study relatively bright 
sources, as discussed in~\cite{bib:2013arXiv1303.2024V}. 
The expected scientific cases shown in the present paper are based on the most updated simulations of a SST--2M mini-array performed 
by the ASTRI Monte Carlo group in synergy with the CTA Monte Carlo group, and whose results are discussed in~\cite{bib:DiPierro_icrc13}. 
The main figure of merit is the sensitivity, expressed as the minimum detectable flux in 50\,hrs of observations. With an energy 
threshold of $\sim 1$\,TeV, the ASTRI mini-array, in a 7-unit layout, has a sensitivity similar or better than that of 
H.E.S.S. in the energy range (a few -- 100)\,TeV (see Figure~6 in \cite{bib:DiPierro_icrc13}).
The other two key parameters, the angular 
and energy resolution, can be as good as 0.08\,deg and 15\% respectively at 10 TeV. Since energies higher than few tens 
of TeV are widely unexplored, dedicated analysis techniques may provide room for further improvement in that regime.

%%%%%%%%%%%%%%%%%%%%%%%%%%%%%%%%%%%%%%%%%%%%%
%%%%%%%%%%%%%%%%%%%%%%%%%%%%%%%%%%%%%%%%%%%%%
	\section{Galactic targets and science}
%%%%%%%%%%%%%%%%%%%%%%%%%%%%%%%%%%%%%%%%%%%%%
%%%%%%%%%%%%%%%%%%%%%%%%%%%%%%%%%%%%%%%%%%%%%
%
The CTA Southern site will provide an excellent view of most of the Galactic plane and of the Galactic Bulge.
Several Galactic sources have been detected so far (see~\cite{bib:2005Sci...307.1938A,bib:2009ARA&A..47..523H}) above a few tens
of TeV, including shell-type supernova remnants (SNR), pulsar wind nebulae (PWN, such as the Crab Nebula), binary systems, 
the Galactic Center, as well as a number of unidentiÞed sources apparently emitting only above 100~GeV
with no lower-energy counterparts.

RX~J1713.7$-$3946 is a young shell-like SNR
which could be considered as an excellent laboratory to investigate the cosmic ray acceleration
(see~\cite{bib:2000A&A...354L..57M} and~\cite{bib:2006A&A...449..223A}).
 \begin{figure}[t]
  \centering
  \includegraphics[width=0.48\textwidth]{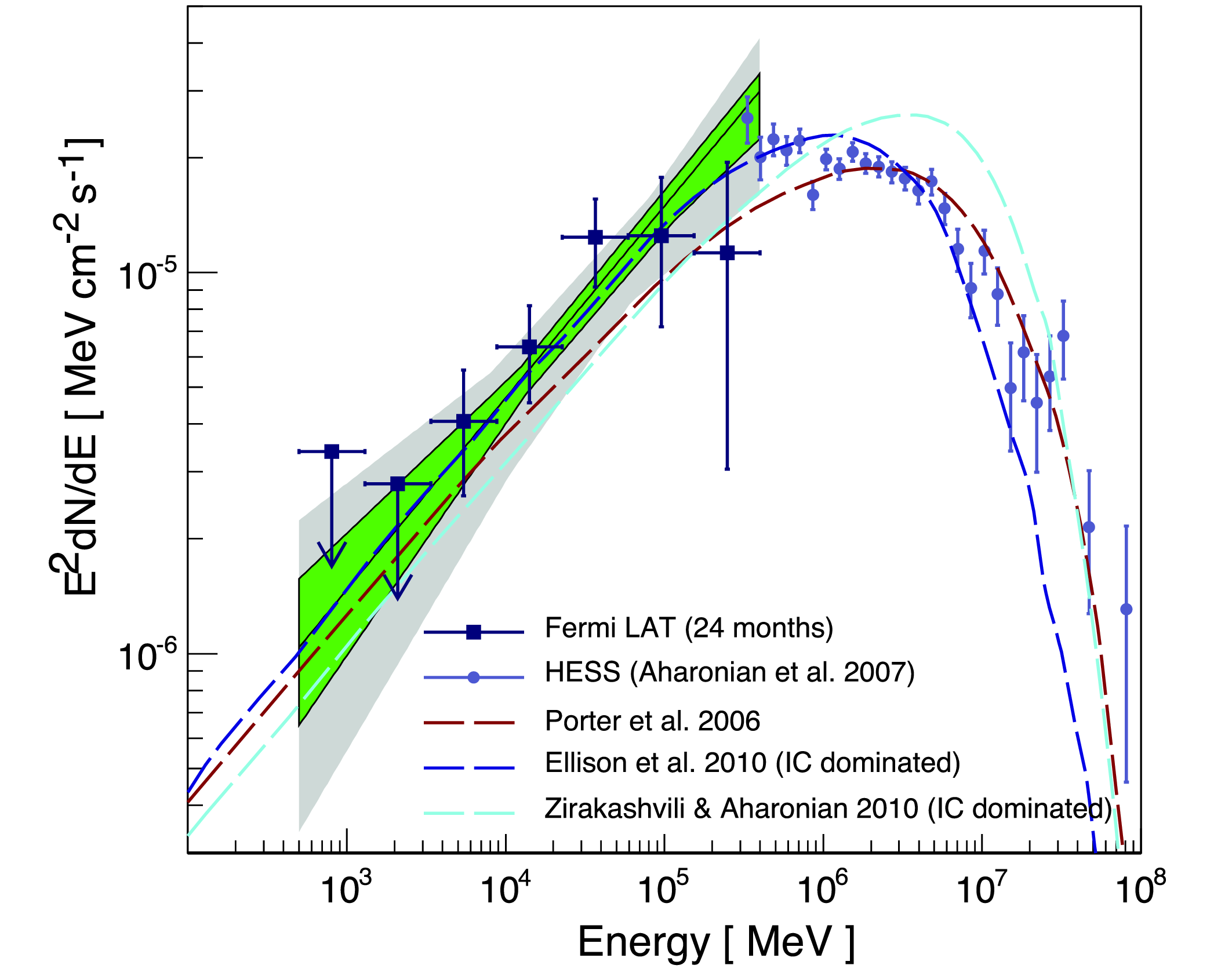}
  \caption{Supernova remnant RX~J1713.7$-$3946. See~\cite{bib:2011ApJ...734...28A} for details.}
  \label{fig:icrc2013_0109_01}
 \end{figure}
The recent detection of this SNR by {\it Fermi} \cite{bib:2011ApJ...734...28A} and the combined study with H.E.S.S.
(see Figure~\ref{fig:icrc2013_0109_01}) show that the high-energy and very high-energy (VHE) emission could be interpreted in the 
framework of a leptonic scenario.
Nev\-er\-the\-less, the majority of current models adopt a spherical geometry, while it is known that the gas distribution around the 
SNR is inhomogeneous~\cite{bib:2013arXiv1304.1261F} (and references therein). A clumpy circum-stellar medium (CSM) 
could produce an hadronic spectrum different  from the prediction of a simple spherical model.  
A detailed comparison between the proton distribution in the CSM and the high-resolution gamma-ray image is therefore 
a useful test of the hadronic scenario.
The improved sensitivity and angular resolution of the mini-array at energy above 10~TeV with respect to the
current IACTs will allow us to investigate the VHE emission in the different regions of this source, studying their spectra.

Tycho's SNR is the best candidate as Pevatron~\cite{bib:2012A&A...538A..81M}, but it will not be accessible from the 
CTA southern site (it is located at $Dec(\rm J2000)\sim +64^{\circ}$ North). Interestingly, the Kepler's SNR, which will be accessible to CTA, 
is very similar to Tycho in many respects and it is expected to produce a similar gamma-ray spectrum.
The H.E.S.S. telescope observed Kepler's SNR for 13~hrs and provided upper limits on the energy flux in the range 
230 GeV -- 12.8 TeV of $8.6 \times 10^{-13}$\,erg\,cm$^{-2}$\,s$^{-1}$~\cite{bib:2008A&A...488..219A}. Indeed, theoretical 
mo\-dels~\cite{bib:2006A&A...452..217B,bib:2012AIPC.1505..241M} predict that the high energy emission from Kepler's SNR should 
be only a factor 2--5 below the H.E.S.S. upper limits. Hence, the mini-array could be able to de\-tect this young SNR
by means of a deep observation, especially if conducted in conjunction with one (or even better \mbox{two}) medium size telescope
units to be placed at the same CTA southern site, which could expand the energy range below a few TeV.

Recently, a lot of interest has been raised on gamma-ray sources produced by middle-age SNRs interacting with 
Molecular Clouds (MC). The detected gamma-ray emission confirmed that hadrons are indeed accelerated in 
\mbox{SNRs}~\cite{bib:2011ApJ...742L..30G,bib:2010Sci...327.1103A}, 
even if the produced spectrum is not the one responsible for the CR flux observed at Earth. In spite of this, such systems 
can provide useful information on how particle escape from the remnant and propagate in their vicinity. Several sources 
like W~28, W~30, W~51C and IC~443 can be observed by the miny-array with better spatial and energy resolution than 
what has been done before.

PWN originate from the interaction of the relativistic wind of a pulsar with the surrounding medium and thus are
excellent candidates for the study, among oth\-ers, of particle acceleration and cooling in relativistic \mbox{shocks}.
From its position on the southern hemisphere, the ASTRI mini-array will be able to observe two of the most notable
examples of TeV-emitting PWN, that is Vela$-$X and \mbox{HESS}~J1825$-$137.
The former is a bright ($\sim75$\% of the Crab flux) and extended ($\sim1$ deg) source, which shows no signs
of spectral softening at increasing distance from its parent's pulsar. This is indicative that some re-acceleration
mechanism is at work in this source~\cite{bib:2012A&A...548A..38A}.
On the other hand, the smaller ($\sim0.2$ deg) and dimmer ($\sim15$\% of
the Crab flux) HESS J1825$-$137 clearly shows such a spectral softening~\cite{bib:2006A&A...460..365A}.
The observations of the mini-array will give stronger constraints on the maximum energy achievable by the
relativistic particles at the termination shock, improving our understanding of the acceleration mechanism and
on the properties of the surrounding medium.

At $\sim1$ deg from HESS~J1825$-$137 lies the microquasar LS~5039. The detection of flux and spectral modulations
in the VHE emission from this source might be indicative of a phase-dependent gamma-ray absorption via pair
production~\cite{bib:2006A&A...460..743A}.
The mini-array will be able to give im\-por\-tant constraints on the spectrum of the observed TeV pho\-tons at phases different
from the superior conjunction (where the softness of the spectrum will be a challenge for the detection capabilities
of the mini-array), thus giving stronger constraints on the gamma-ray emission and absorption.

%%%%%%%%%%%%%%%%%%%%%%%%%%%%%%%%%%%%%%%%%%%%%
%%%%%%%%%%%%%%%%%%%%%%%%%%%%%%%%%%%%%%%%%%%%%
	\section{Extra-galactic targets and science}
\begin{figure}[t]
  \centering
  \includegraphics[width=.48\textwidth]{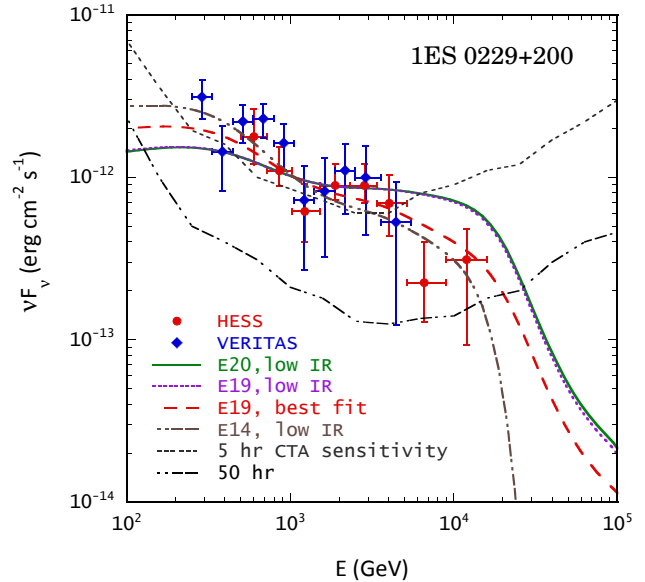}
  \caption{Spectral energy distribution of the extreme blazar 1ES~0229$+$200 (see~\cite{bib:2012ApJ...749...63M} for details).
  We expect for the SST--2M mini-array in a 7-unit layout a sensitivity at least comparable to (or slightly better than) the 
  H.E.S.S. one above a few TeV.
  }
  \label{fig:icrc2013_0109_02}
\end{figure}

%%%%%%%%%%%%%%%%%%%%%%%%%%%%%%%%%%%%%%%%%%%%%
%%%%%%%%%%%%%%%%%%%%%%%%%%%%%%%%%%%%%%%%%%%%%
%
The ASTRI mini-array will be extremely important to investigate the VHE emission from extragalactic sources as well.
Figure~\ref{fig:icrc2013_0109_02} shows the spectral energy distribution (SED) of the extreme blazar (E-HBL) 
1ES~0229$+$200 \cite{bib:2012ApJ...749...63M}
with superimposed the cascade spectra (E20 low IR and E19 low IR) initiated by the $E^{-2}$ injection with 
$E^{\rm max}_{\rm p} = 10^{20}$\,eV and $10^{19}$\,eV protons, respectively, using the low-IR EBL model 
by~\cite{bib:2004A&A...413..807K},  while the red-dashed curve (E19 best-fit) is the spectrum with 
$E^{\rm max}_{\rm p} = 10^{19}$\,eV the best-fit EBL model. The brown dot-dot, dash-dash curve
(E14 low IR) is the spectrum resulting from the cascade of $E^{\rm max}_{\rm p} = 10^{14}$\,eV
primary photons with index $\beta = 5/4$ produced at the source for the low-IR EBL model. Double
dot-dashed and dotted curves show, respectively, the 5-$\sigma$ differential sensitivity
for 5 and 50\,hr observations with CTA (configuration E, as reported in~\cite{bib:2011ExA....32..193A}).
As discussed in~\cite{bib:2012ApJ...749...63M} a clear detection of VHE emission above a few tens of TeV from such a blazar 
could provide a striking evidence for non-standard phenomena, either an anomalous transparency of the Universe at these 
energies~\cite{bib:2012JCAP...02..033H,bib:2013arXiv1302.6460D} or gamma-ray emission resulting from an electromagnetic cascade 
initiated by ultra-relativistic protons accelerated in the blazar jet and beamed toward the observer. A list of promising E-HBLs is currently under 
study ~\cite{bib:Bonnoli_inprep} and will provide a fundamental ``target list'' for the future mini-array.
The latter possibility would also demonstrate the possibility that relativistic jets are the accelerators 
of the still enigmatic UHECR.

\begin{figure}[t]
  \centering
  \includegraphics[width=.48\textwidth]{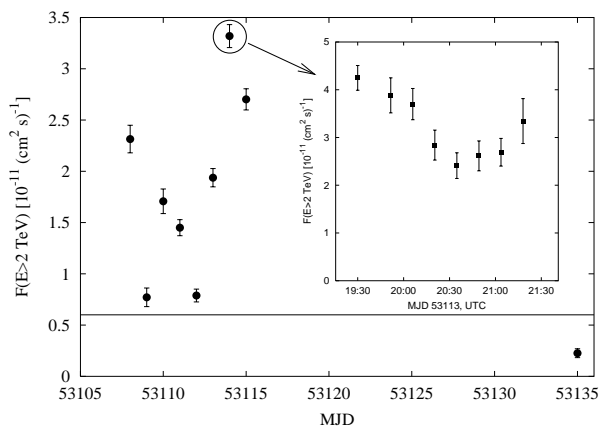}
  \caption{MKN~421 daily light-curve at energy $E>2$\,TeV. The inset shows possible intra-night variability (time-bin of 14\,min) during
  a particularly high flux-state (see~\cite{bib:2005A&A...437...95A} for details).
  }
  \label{fig:icrc2013_0109_03}
\end{figure}

The study of the extra-galactic background light (\mbox{EBL}) in the far infrared energy band could be an important task, although not
a simple one, for the mini-array. Possible candidates should be nearby, hard, intense blazars. Among those observable
form the Southern Hemisphere, we can consider MKN~421 ($z=0.03$) and possibly M~87 ($z=0.0043$), the latter one being
less intense than the former. MKN~421 and M~87 will be observable from the \mbox{CTA} southern site at high zenith-angles, requiring
ad-hoc Monte Carlo simulations in order to fully study their VHE properties. If detected above 10\,TeV, M~87 would be particularly
important to investigate the VHE emission mechanisms in radio-galaxies. 
Observations of MKN~421 above 10~TeV could be crucial, especially during high or very-high states, not only
for the EBL studies, but also for the intrinsic relevance of this source.
These ob\-ser\-va\-tions will allow us to investigate the intra-night variability of such intense and low-redshift class of
objects above a few TeV~\cite{bib:2005A&A...437...95A} during periods of high-flux states, as shown
in Figure~\ref{fig:icrc2013_0109_03}.
%

%%%%%%%%%%%%%%%%%%%%%%%%%%%%%%%%%%%%%%%%%%%%%
%%%%%%%%%%%%%%%%%%%%%%%%%%%%%%%%%%%%%%%%%%%%%
	\section{The SST--2M mini-array wide field of view}
%%%%%%%%%%%%%%%%%%%%%%%%%%%%%%%%%%%%%%%%%%%%%
%%%%%%%%%%%%%%%%%%%%%%%%%%%%%%%%%%%%%%%%%%%%%
%
\begin{figure}[t]
  \centering
  \includegraphics[angle=90,width=.48\textwidth]{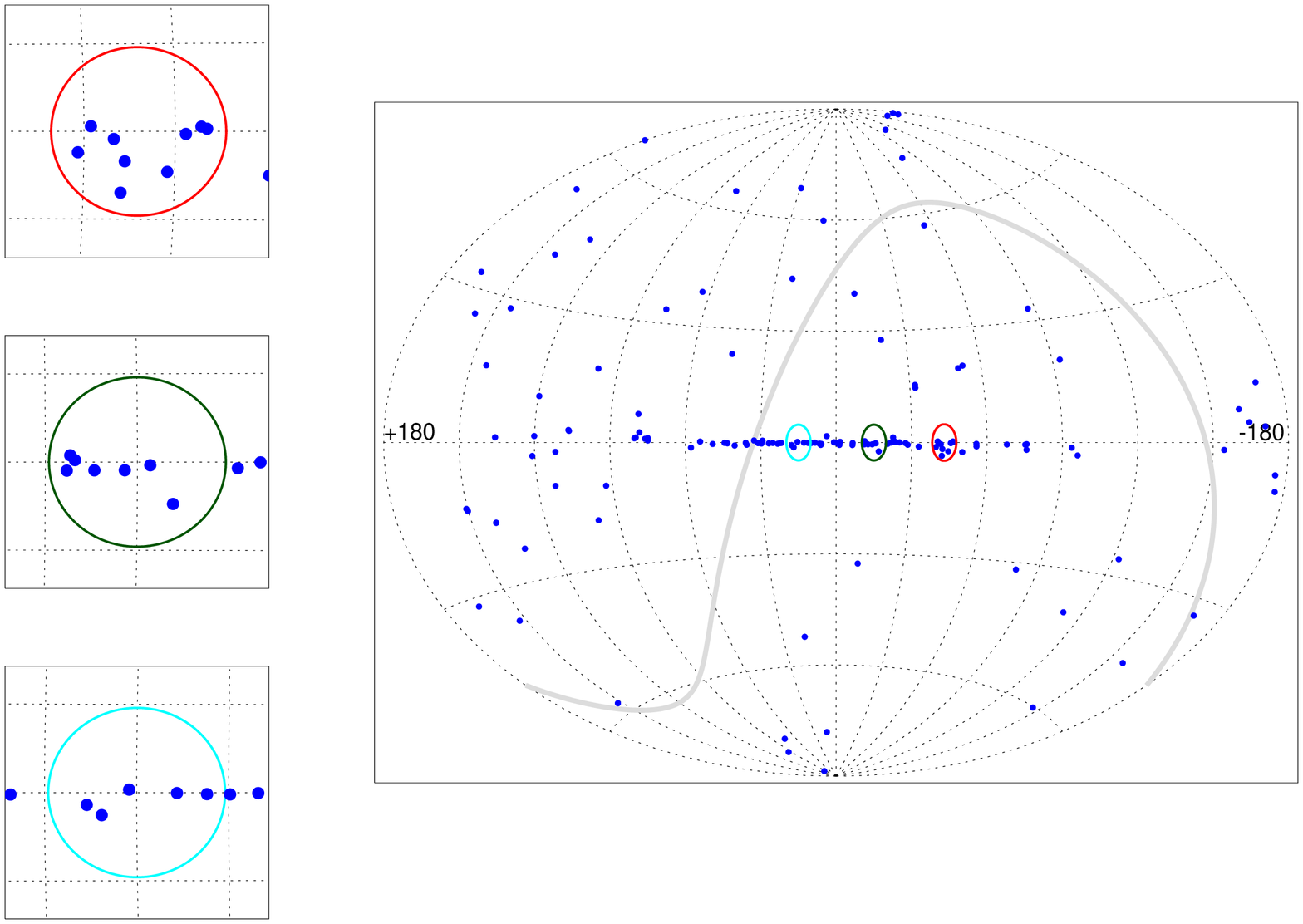}
  \caption{Blue dots are the known TeV sources as listed in the TeVCat Catalogue. The grey line represents the
  Celestial Equator. The red, green and cyan circles are the ASTRI mini-array (optical) field of view during three possible
  pointings. The left panels are zooms centered on the ASTRI mini-array pointings. Different colors are for sake of clarity only.
  }
  \label{fig:icrc2013_0109_04}
\end{figure}
The large field of view of the ASTRI mini-array will allow us to monitor, during a single pointing, a few TeV sources
simultaneously. Figure~\ref{fig:icrc2013_0109_04} shows the current TeV sources as listed in the 
TeVCat\footnote{\texttt{http://tevcat.uchicago.edu/}} compilation. Red, green and cyan circles represent the $9.6^{\circ}$ 
(optical) field of view diameter for three possible pointings along the Galactic Plane. 
The grey line represents the Celestial equator.
Although the actual sensitivity will substantially drop for off-axis sources, a few targets can be monitored simultaneously, as 
shown in the three panels on the left. Simultaneous detection of hard and intense Galactic sources could be feasible,
e.g. in the case of Vela--X and Vela--Jr. Moreover, detections of serendipitous strong flares (a few Crab units) from hard-spectrum 
sources will be possible as well. We also notice that several GeV sources detected by {\it Fermi}-LAT lie 
within the central region of each mini-array pointings.
%

%%%%%%%%%%%%%%%%%%%%%%%%%%%%%%%%%%%%%%%%%%%%%
%%%%%%%%%%%%%%%%%%%%%%%%%%%%%%%%%%%%%%%%%%%%%
	\section{Conclusion}
%%%%%%%%%%%%%%%%%%%%%%%%%%%%%%%%%%%%%%%%%%%%%
%%%%%%%%%%%%%%%%%%%%%%%%%%%%%%%%%%%%%%%%%%%%%
%
The SST--2M mini-array, whose operations are planned to start since 2016, could constitute 
the first {\it seed} of the future CTA Project, and will be open to the CTA Consortium for both 
technological and scientific exploitation.
A robust improvement could come from the implementation of one (or even better two) medium-size telescope
at the same site. This would extend the energy range down to a few hundreds of GeV, lower the trigger
threshold, and largely improve the science return.

\vspace*{0.5cm}
\footnotesize{{\bf Acknowledgment:}{This work was partially supported by the ASTRI Flagship Project financed by the Italian 
Ministry of Education, University, and Research (MIUR) and led by the Italian National Institute of Astrophysics (INAF). 
We also acknowledge partial support by the MIUR Bando PRIN 2009.}}

\end{document}